\documentclass[reprint,prl,aps,floatfix,amsmath,amssymb,superscriptaddress]{revtex4-2}
\usepackage[T1]{fontenc}
\usepackage{bm,graphicx}
\usepackage{xcolor}

\usepackage[colorlinks=true,citecolor=blue,urlcolor=blue,linkcolor=red]{hyperref}
\newcommand{\hv}{\hbar v_F}
\newcommand{\norm}[1]{\lVert #1\rVert}
\newcommand{\diag}{\operatorname{diag}}
\usepackage{placeins}

\begin{document}

\title{Topological Bound States in the Continuum}
\author{Dan-Na Liu}\email{dannaliu@hainanu.edu.cn}
\affiliation{Center for Theoretical Physics, Hainan University, Haikou 570228, China}
\affiliation{School of Physics and Optoelectronic Engineering, Hainan University, Haikou 570228, China}
\author{J. Villavicencio}
\affiliation{Facultad de Ciencias, Universidad Aut\'onoma de Baja California, Carretera Transpeninsular Tijuana-Ensenada No.~3917, Ensenada, 22800, B.C., M\'exico}
\author{Pierre A. Pantale\'on}
\affiliation{Center for Theoretical Physics, Hainan University, Haikou 570228, China}
\affiliation{School of Physics and Optoelectronic Engineering, Hainan University, Haikou 570228, China}

\begin{abstract}
We uncover topological bound states in the continuum in
AB-stacked bilayer graphene mass-wall junctions.\ Valley topology
supports interface states propagating along the wall, while a
distinct winding of the confinement conditions governs exact
decoupling from conducting leads.\ Tuning the junction length or
interlayer bias can merge oppositely charged bound states and
remove confinement without changing the valley topology.\ At
generic critical points, the state remains bound and nearby
resonance widths open quartically with momentum.\ These results
connect complementary roles of topology in interface propagation
and confinement within a conducting continuum.\
\end{abstract}
\maketitle

Confinement in an open system does not necessarily require a spectral
gap.\ Bound states in the continuum (BICs), first proposed by von Neumann
and Wigner~\cite{VonNeumannWigner1929}, remain localized despite
coexisting with extended states, owing to symmetry or destructive
interference~\cite{FriedrichWintgen1985,Hsu2016}.\ Electronic BICs have
been observed in semiconductor heterostructures~\cite{Capasso1992} and
studied in quantum dots and waveguides, where they can be associated
with the collapse of transmission resonances~\cite{KimJETP1999,Ordonez2006}.\
Recently, scanning tunneling microscopy identified defect-induced
electronic states within the metallic continuum of
$\mathrm{Pd}_5\mathrm{AlI}_2$, associated with hopping interference
and a nearly flat band~\cite{Thinel2026}.

Topology provides a connection between the motion of localized
states and their confinement in an open environment.\ Experiments
in acoustic moir\'e superlattices have demonstrated pumping of
localized bulk states governed by a band Chern
number~\cite{Yao2026Directional}.\ Topological protection against
hybridization with a continuum has been predicted for a quantum
Hall layer coupled to a topologically trivial
substrate~\cite{Yang2013Hybridization}, and topological BICs have
been proposed in electronic multilayer models~\cite{Takeichi2019}.\
In acoustics, valley interface states have been realized that
propagate while remaining bound within a bulk
continuum~\cite{Yin2024ValleyBIC}.\ A distinct topological description
arises from the coupling to extended states: in photonic crystal
slabs, radiation amplitudes form momentum-space vortices whose
integer charges constrain the evolution and annihilation of
BICs~\cite{Zhen2014}.\ Their merging can suppress radiative losses
and increase the lifetimes of nearby resonances~\cite{Jin2019},
connecting the topology of decoupling to measurable escape rates.\
Related scattering singularities have also been studied in open
wave systems~\cite{Sakotic2023}.

Graphene provides an electronic setting in which several mechanisms
of continuum decoupling can be explored.\ BICs have been predicted
in quantum dots~\cite{Gonzalez2010}, trilayer flakes connected to
nanoribbon leads~\cite{Cortes2014}, and defect-containing bilayer
materials, including graphene, where confinement follows from
the algebraic structure of the minimal Hamiltonian~\cite{Massatt2025}.\
In bilayer graphene, a different mechanism arises from chirality,
which can suppress coupling between confined and propagating
states and produce electronic cloaking~\cite{Gu2011}, with
experimental evidence reported in Ref.~\cite{LeeCloaking2016}.\
Mode-resolved barrier calculations clarify how propagating and
evanescent channels, symmetry, and phase matching determine
this coupling~\cite{Nilsson2007Transmission,van2013four,LiuCloaking2026}.

Interlayer bias allows a complementary confining structure.\ In AB-stacked bilayer graphene, an interlayer
potential difference opens a gap~\cite{McCann2006}, while junctions with sign reversal create a mass wall supporting one-dimensional interface
states~\cite{Martin2008,Zhang2013PNAS}.\ These states are localized
across the wall and propagate along it, with their existence
related to the difference in valley Chern numbers between the
massive domains~\cite{Martin2008}.\ Gate-defined channels have been observed experimentally~\cite{Li2016}
and used to realize valley valves and electron beam
splitters~\cite{Li2018}.\ Gate-controlled valley polarizers have also
demonstrated valley-selective transport at zero magnetic
field~\cite{Chen2020ValleyPolarizer}, while crossed gates provide
reconfigurable point junctions carrying valley-selective
current~\cite{Davydov2024PointJunction}.\ Mass walls therefore
provide an electrically controlled platform for examining the
relation between topological interface propagation and confinement.

Connecting a finite mass-reversal region to conducting leads
allows its interface states to couple to the continuum through
their transverse tails. This coupling can give the states a
finite lifetime and produce resonances in transport across the
junction, even while the mass reversal and valley topology
remain intact. Valley topology alone therefore does not
determine their confinement in the open junction.

In this Letter, we establish exact confinement of mass-wall states
in bilayer graphene through a general matching condition,
accompanied by the collapse of transmission resonances.
In symmetric junctions, this condition separates into mirror
sectors, whose isolated zeros carry integer winding charges.
Length or mass tuning can bring opposite charges together and
remove exact confinement, with quartic linewidth opening at the
critical point for regular continuum coupling.
Exact confinement can also persist for unequal masses through
additional tuning.
These results connect the valley topology supporting interface
propagation with a distinct topology governing confinement and
escape into conducting leads, providing a theoretical basis
for exploring electrically tunable confinement in gate-defined
valley devices.

\begin{figure*}
\includegraphics[width=0.95\linewidth]{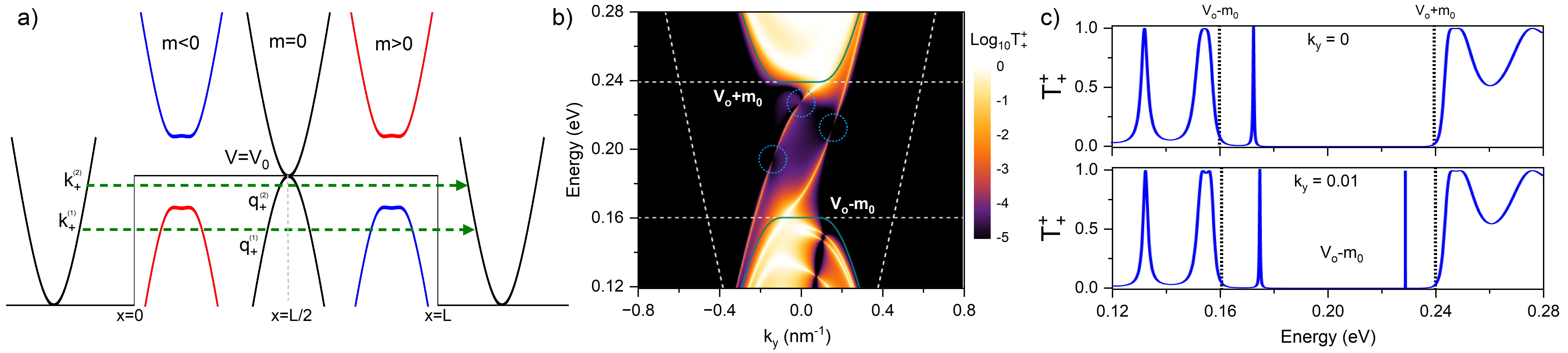}
\caption{{\bf Topological confinement and resonance collapse}.
(a) Schematic of the junction and its local band structure; green dashed
arrows indicate two incident energies.
(b) Transmission probability $T(E,k_y)$ in the channel $k_p$ on a
logarithmic scale. Circles mark the three BIC positions identified by decaying-tail matching. White dashed curves delimit the propagating lead channel, and solid green curves indicate the projected band edges of the massive domains. The horizontal dashed lines indicate $V_0\pm m_0$. (c) Energy cuts at $k_y=0$ (upper panel) and $k_y=0.01\,\mathrm{nm}^{-1}$ (lower panel). Narrow resonances occur in the projected gap. The resonance near $E=0.23\,\mathrm{eV}$, visible at finite $k_y$, disappears at normal incidence. Its collapse is accompanied by an exact decaying solution in Eq.~\eqref{eq:map}. Panels (b,c) use a sharp junction at $d\simeq23.3$ nm.}
\label{fig:transport}
\end{figure*}

\textit{The model.---}We consider a bilayer graphene junction invariant
along $y$, with $\Psi(x,y)=e^{ik_y y}\Phi(x)$. As shown in Fig.~\ref{fig:transport}a), the region $0<x<L$ is embedded between pristine, semi-infinite bilayer graphene leads. In the basis $(A_1,B_1,B_2,A_2)$ and for one valley, the minimal
four-band Hamiltonian is
\begin{equation}
H_{k_y}=\begin{pmatrix}
V+m & \hv\pi & \gamma_1 & 0\\
\hv\pi^\dagger & V+m & 0 & 0\\
\gamma_1 & 0 & V-m & \hv\pi^\dagger\\
0 & 0 & \hv\pi & V-m
\end{pmatrix},
\label{eq:H}
\end{equation}
where $\pi=-i\partial_x+ik_y$, $\gamma_1$ is the interlayer hopping,
$v_F$ is the monolayer Fermi velocity, $2m$ is the layer-potential
difference, and $V$ a scalar potential. We set $V=m=0$ in
the leads and $V=V_0$ inside the junction. A sharp wall at $x=d_L$
separates domains with masses $-m_L$ and $+m_R$, where $m_L,m_R>0$,
and widths $d_L$ and $d_R=L-d_L$. The symmetric junction in
Fig.~\ref{fig:transport}(a) has $m_L=m_R=m_0$ and
$d_L=d_R=d=L/2$. We work in the energy and transverse-momentum range in which each lead supports one propagating and one evanescent channel, given by
\begin{equation}
0<E<\gamma_1,\qquad
k_{\rm p}^2=\frac{E(E+\gamma_1)}{(\hv)^2}-k_y^2>0.
\label{eq:continuum}
\end{equation}
The lead modes and their decay constant $\kappa>0$ are given in Sec.~\ref{supp:sec:model} of
the Supplemental Material. Continuity of all four components
of $\Phi$ determines the scattering amplitudes~\cite{van2013four}.\
For identical leads the transmission probability is $T=|t|^2$.
Unless stated otherwise, we use $\hv=0.6582\,\mathrm{eV\,nm}$,
$\gamma_1=0.4\,\mathrm{eV}$, $V_0=0.2\,\mathrm{eV}$, and
$m_0=0.04\,\mathrm{eV}$ to illustrate our approach numerically.\
We note that the behavior discussed below also occurs for other
parameter choices.

For the symmetric wall, the two massive domains have a common bulk
gap extending from $V_0-\Delta_{\rm bulk}$ to $V_0+\Delta_{\rm bulk}$,
with $\Delta_{\rm bulk}=m_0\gamma_1/\sqrt{\gamma_1^2+4m_0^2}$.
The mass reversal supports interface states within this
gap~\cite{Martin2008}. Interestingly, we found that their coupling to the leads produces the
dispersive resonances in Fig.~\ref{fig:transport}(b), which connect
the lower and upper regions of the spectrum. Although the states
propagate along the wall, this coupling makes them visible in
transport across the junction. The circled suppressions and the
energy cuts in Fig.~\ref{fig:transport}(c) show that some resonances
disappear at selected $(E,k_y)$. This raises the question of whether the corresponding wall states remain bound even though the leads support propagating modes.

A similar resonance collapse occurs in scalar barriers at normal
incidence, where bilayer chirality can decouple confined states
from propagating carriers~\cite{Katsnelson2006Chiral,Gu2011}.
In the present junction, however, the origin is distinct since the mass modifies the coupling between
modes~\cite{Nilsson2007Transmission}, while its reversal supports
states localized near the wall. The evanescent tails of the states in the wall extend through the massive domains and
can reach the leads, allowing escape even within the bulk gap.\

\textit{Bound states in the continuum.---} We now analyze the conditions for this effect to occur.\
At fixed $k_y$, we seek a nonzero eigenstate with finite norm
in $x$ and dependence $e^{ik_y y}$ along the wall.\
A finite norm excludes propagating components in the leads,
leaving only the decaying modes, with spatial factors
$e^{\kappa x}$ on the left and $e^{-\kappa(x-L)}$ on the right.\
Denoting their contact spinors by $\bm u_L$ and $\bm u_R$,
we propagate them to a common point $x_*$ inside the junction
to determine whether they can be joined into one continuous
eigenfunction.
Let $\bm v_L$ and $\bm v_R$ denote the resulting spinors, with
the right tail propagated backward. They match continuously
through the junction precisely when
\begin{equation}
\mathcal D_{\rm BIC}
=\norm{\bm v_L}^{2}\norm{\bm v_R}^{2}
-|\bm v_L^\dagger\bm v_R|^2=0.
\label{eq:general}
\end{equation}
Together with Eq.~\eqref{eq:continuum}, this condition establishes
a BIC: the two tails join into a nonzero eigenfunction with finite
norm in $x$. The result is independent of the matching point and
tail normalization. Internal propagating components are allowed,
and the same construction applies to nonuniform profiles using
finite slabs~\cite{LiuMode2026}. The propagation and matching
conditions are given in Sec.~\ref{supp:sec:matching} of Supplementary Material.

In our system, the bilayer Hamiltonian has the antiunitary symmetry
$\Theta=Z\mathcal K$, with $Z=\operatorname{diag}(1,-1,1,-1)$
and $\mathcal K$ complex conjugation. It satisfies $\Theta^2=1$
within one valley at fixed $k_y$, whereas physical time reversal
exchanges valleys and reverses momentum. The lead spinors obey
$Z\bm u_{L,R}^*=\bm u_{L,R}$, and propagation preserves this
relation. Their propagated $A_1,B_2$ components are therefore
real and their $B_1,A_2$ components imaginary in this normalization. Matching requires $\bm v_L=\alpha\bm v_R$, where $\alpha$ is
a real amplitude ratio in this normalization.\ A component equation with a nonzero component of $\bm v_R$
determines $\alpha$; the remaining three component equations
must then hold with the same ratio.\ Equation~\eqref{eq:general} therefore collects three real matching
conditions into a single scalar equation (for details see Sec.~\ref{supp:sec:matching} in Supplementary Material).

\textit{Exact mirror reduction.---}The matching problem simplifies
when reflection relates the two halves of the junction. For
$V(L-x)=V(x)$ and $m(L-x)=-m(x)$, reflection about the wall combined
with an exchange of layer and sublattice components leaves the
Hamiltonian unchanged. The resulting mirror operator satisfies
$\mathcal M^2=1$, allowing a BIC to be assigned a parity
$\sigma=\pm1$. At $d=L/2$, this requires
$\Phi_{A_1}(d)=\sigma\Phi_{B_2}(d)$ and
$\Phi_{B_1}(d)=-\sigma\Phi_{A_2}(d)$.

These relations let us determine confinement from the left half.
Writing $P_L=\mathsf P(d,0)$ for propagation from the left contact
to the wall, the sector condition is
\begin{equation}
\bm r_\sigma(E,k_y;d)
=\mathsf B_\sigma P_L(E,k_y;d)\bm u_L(E,k_y)=0,
\label{eq:sector}
\end{equation}
where $\mathsf B_\sigma$ imposes the two component relations above.
The right half follows by reflection with parity $\sigma$ and
matches the decaying mode in the right lead. The explicit symmetry
and boundary matrices are given in Sec.~\ref{supp:sec:matching} of Supplementary Material.
Each mirror sector also contains a propagating channel formed
from incidence in the two leads. This differs from the acoustic
valley BICs of Ref.~\cite{Yin2024ValleyBIC}, where separation
into independent subsystems prevents an interface state in one
subsystem from hybridizing with the continuum of the other.
Here, the BIC and propagating lead states share the same mirror
parity. Exact confinement therefore requires the additional
matching condition in Eq.~\eqref{eq:sector} to cancel their
coupling within that sector.

The antiunitary symmetry makes $r_{\sigma,1}$ real and
$r_{\sigma,2}$ imaginary.\ We denote the two real matching
conditions by $F_{\sigma,1}=\operatorname{Re}r_{\sigma,1}$ and
$F_{\sigma,2}=\operatorname{Im}r_{\sigma,2}$, collected in
$\bm F_\sigma=(F_{\sigma,1},F_{\sigma,2})^T$. At the center,
the general condition factorizes exactly as
\begin{equation}
f_\sigma=F_{\sigma,1}+iF_{\sigma,2},\qquad
\mathcal D_{\rm BIC}=|f_+|^2|f_-|^2.
\label{eq:map}
\end{equation}
Thus $f_\sigma=0$ identifies a BIC in sector $\sigma$.\ The factorization follows from reflection of the propagated tails,
as shown in Sec.~\ref{supp:sec:matching} of the Supplementary Material.

\textit{Vortex--antivortex annihilation.---}The complex matching
function also describes how BICs evolve under changes in the
junction parameters. Around an isolated zero, its phase can wind
by an integer multiple of $2\pi$. For a counterclockwise contour
$\mathcal C$ in the ordered $(E,k_y)$ plane, with $f_\sigma\ne0$
on the contour, the winding charge is
\begin{equation}
q_\sigma=\frac{1}{2\pi}\oint_{\mathcal C}d\arg f_\sigma.
\label{eq:winding}
\end{equation}
A simple zero carries charge $q_\sigma=\pm1$. Under smooth changes
preserving the mirror and antiunitary symmetries and the lead-channel
structure, the enclosed winding remains unchanged unless a zero
crosses $\mathcal C$. A nonzero winding therefore guarantees a
matching zero inside the contour, although its position can move.
The corresponding local derivative formula is given in Sec.~\ref{supp:sec:topology} of Supplementary Material.

Figure~\ref{fig:matching_maps} shows these zeros through the phase
$\arg f_\sigma$ and the direction of $\bm F_\sigma$.\ The energy--momentum maps contain isolated BICs in both sectors,
including normal- and oblique-incidence solutions. The
energy--length and energy--mass sections identify selected
combinations of energy and half-length $d$ or mass magnitude
$m_0$ at which leakage vanishes. The phase is undefined at each
marked zero and winds around it, giving to the exact continuum decoupling
a topological characterization.

\begin{figure}
\includegraphics[width=0.99\linewidth]{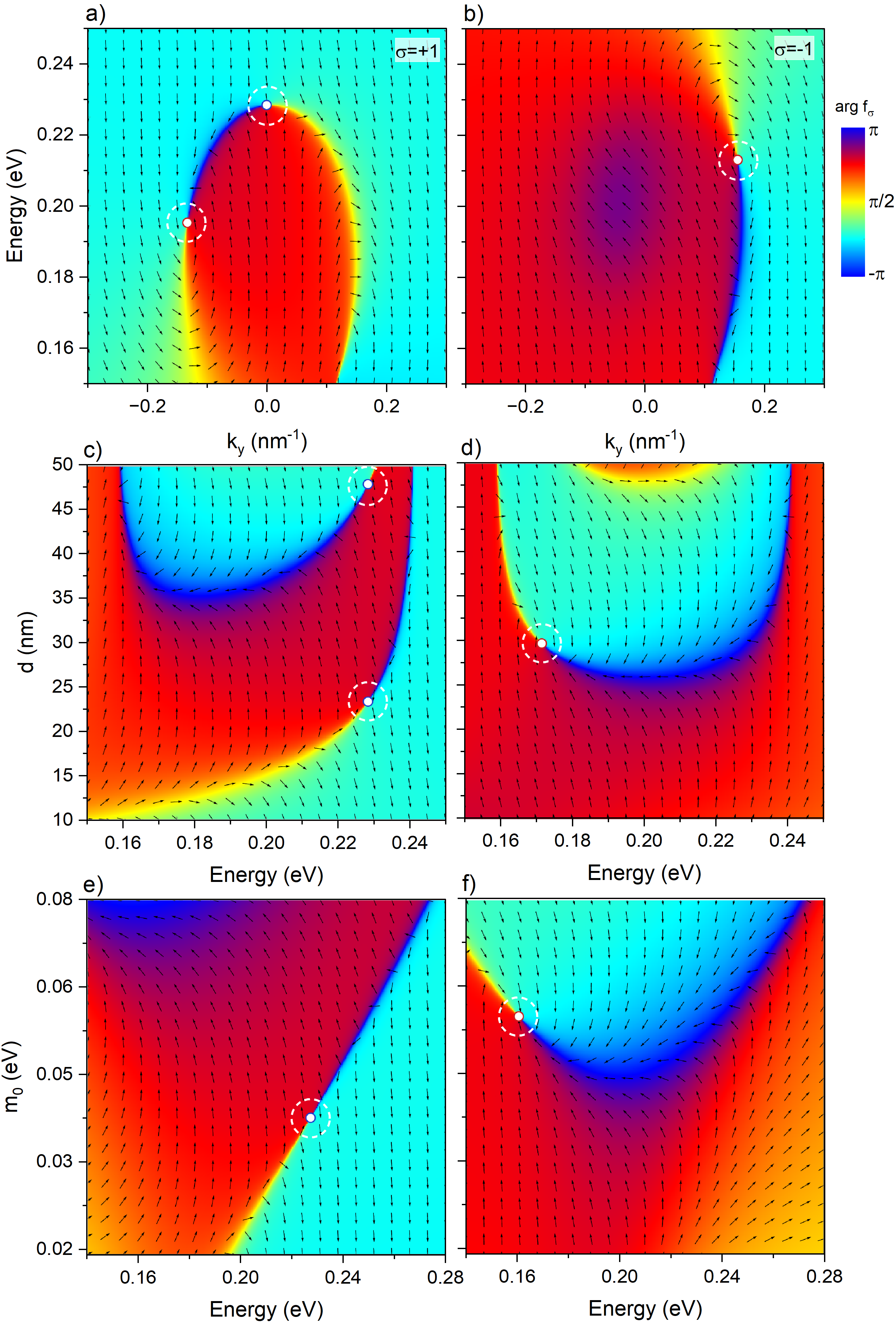}
\caption{{\bf Sector-resolved BICs and phase winding}. The left column (a,c,e) shows the mirror sector $\sigma=+1$, and the right column (b,d,f) shows $\sigma=-1$. Colors represent $\arg f_\sigma$, where  $f_\sigma=F_{\sigma,1}+iF_{\sigma,2}$ is constructed from the two real matching conditions, and arrows indicate the normalized matching vector $\mathbf F_\sigma/|\mathbf F_\sigma|$.
Panels (a,b) show the $(k_y,E)$ plane at $d\simeq23.3$ nm
and $m_0=0.04$ eV; (c,d) show the $(E,d)$ plane at $k_y=0$ and
$m_0=0.04$ eV; and (e,f) show the $(E,m_0)$ plane at $k_y=0$
and $d\simeq23.3$ nm.
Here, $d$ is the half-length of the junction, $L=2d$.
White-filled markers show the BICs satisfying $f_\sigma=0$. White dashed circles identify their position for better visualization.
Blue and red marker borders denote winding numbers $-1$ and $+1$,
respectively. In (a,b), these signs are opposite
to $q_\sigma$ defined in the ordered $(E,k_y)$ plane in
Eq.~\eqref{eq:winding}; the other rows refer to windings in their
respective parameter sections.}
\label{fig:matching_maps}
\end{figure}

Both $d$ and $m_0$ preserve mirror
symmetry. As either changes, simple BIC zeros trace curves in
$(E,k_y,d)$ at fixed $m_0$, or in $(E,k_y,m_0)$ at fixed $d$;
their intersections with $k_y=0$ give normal-incidence solutions.
Two zeros can approach and coincide at a critical length or mass.
For the pair to disappear, its total winding must vanish, so the
two charges must be opposite. Equal charges cannot annihilate
together, while opposite charges alone do not guarantee that a
chosen parameter variation brings them together. Reversing a
pair-annihilation process creates the pair. This evolution has
the same local topological structure as the merging of oppositely
charged Dirac points observed in a tunable honeycomb optical
lattice~\cite{Tarruell2012Creating}.

We found that winding is conserved separately in each mirror sector.\ Charges
belonging to $f_+$ and $f_-$ cannot cancel. Moreover, the unique
decaying lead mode cannot satisfy both mirror-sector boundary
conditions, so opposite-parity BICs cannot coincide at the same
$(E,k_y)$ for a fixed junction. At the critical point where a pair
within one sector coincides, the matching condition still holds
and the state remains a BIC. On the side without the pair, no exact
matching solution exists locally, but electronic weight can remain
concentrated near the wall as a resonance with a finite escape rate.

Remarkably, this matching topology complements the valley topology of the
massive domains. When intervalley scattering is negligible, the
change in valley Chern number across the wall supports states
localized across it and propagating along it~\cite{Martin2008,Zhang2013PNAS}. In the open junction, exact confinement additionally requires their coupling to the conducting leads to vanish. Annihilation of matching zeros can therefore remove exact confinement while the bulk valley indices remain unchanged and a resonance associated with the
interface state survives.

\begin{figure*}
\includegraphics[width=0.99\linewidth]{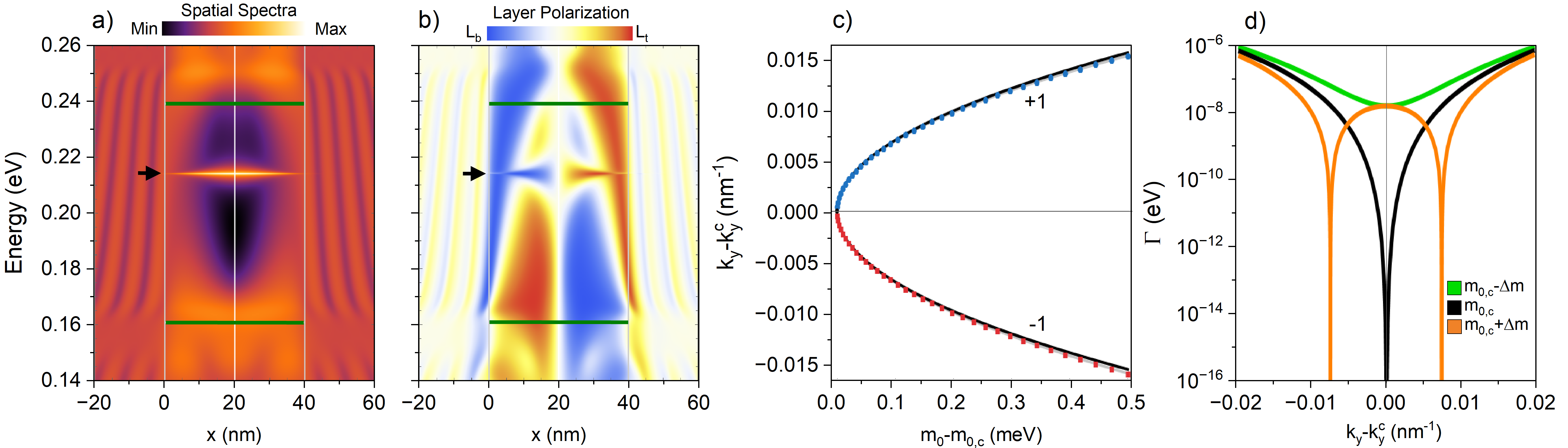}
\caption{{\bf Spatial localization, BIC coalescence, and linewidth opening.} Illustrative mass-driven critical point within one mirror sector
at fixed $d=d_c\simeq20.7$ nm, with $m_{0,c}=40$ meV and $\sigma=+1$. (a) Spatial spectral density and (b) layer polarization at the critical momentum ($k_{y,c}\simeq-0.066\,\mathrm{nm}^{-1}$), evaluated with spectral broadening $\eta=0.10$ meV. Vertical white lines mark the contacts and mass wall; green lines indicate the bulk gap edges. Arrows identify the BIC. In (c) two BICs with opposite winding charges ($q_\sigma=+1$ and $q_\sigma=-1$) approach and merge as the mass magnitude reaches its critical value. Blue circled and red squared lines denote charges $+1$ and $-1$, respectively. Black curves show the local square-root prediction. (d) Resonance linewidth along its dispersion for $\Delta m=0.001$. The merged state remains a BIC, with quartic linewidth opening. Above the critical mass, two BICs exhibit quadratic opening; below it, the local minimum is finite. Zero linewidth lies outside the logarithmic display range. Length tuning at fixed mass produces analogous coalescence.}
\label{fig:coalescence}
\end{figure*}

\textit{Spatial localization and layer polarization.---} The matching
condition identifies exact confinement, while spatial spectra show
where the electronic weight lies and how it is distributed between
the layers.\ Figure~\ref{fig:coalescence}(a,b) shows the local spectral
density defined as $D_G^\eta=D_{1G}^\eta+D_{2G}^\eta$ with a layer polarization given by $P_L^\eta=(D_{1G}^\eta-D_{2G}^\eta)/D_G^\eta$.\ These expressions are given in terms of the  retarded Green function at $E+i\eta$, with $\eta>0$ (Sec.~\ref{supp:sec:spectra} of Supplementary Material).\ These observables include both the BIC and the continuum background. Separating the normalized BIC pole from the remainder of the Green function
gives their additive contributions to each layer density.

The BIC matches only the decaying lead mode, whereas continuum
states also contain propagating components. These boundary
conditions select different component amplitudes and relative
phases. At $k_y=0$, for example, the BIC tail satisfies
$\Phi_{B_2}(0)=\Phi_{A_1}(0)$ and
$\Phi_{A_2}(0)=\Phi_{B_1}(0)$, while the pure propagating channel
has the corresponding minus signs. Propagation through the
massive domains converts these different mode combinations
into different local layer distributions.

The total layer polarization in Fig.~\ref{fig:coalescence}(b)
reverses near the BIC. At the boundary between the two signs,
$D_{1G}^\eta=D_{2G}^\eta$ at finite total spectral density.
The spectral decomposition in Sec.~\ref{supp:sec:spectra} of the Supplemental
Material explains how this reversal can occur. If the
BIC adds more spectral weight to the layer with less continuum
weight, it reverses the local polarization once its layer
imbalance exceeds that of the continuum. Mirror symmetry exchanges
the layers between $x$ and $L-x$, reversing the polarization
of each contribution. The relative sign of the BIC and continuum
polarizations at a given $x$ depends on their local layer weights.
The resulting contrast shows how the BIC changes the layer
character of the local spectrum. Exact confinement is established
by the vanishing of the propagating amplitudes at both contacts,
as expressed by Eq.~\eqref{eq:general}. The corresponding flux
condition is given in Sec.~\ref{supp:sec:topology} of the Supplemental Material.

\textit{Coalescence and linewidth scaling.---} The state in Fig.~\ref{fig:coalescence}(a,b) is the BIC formed when the two solutions in panel (c) merge. When $\partial_E F_{\sigma,1}\ne0$, we solve
$F_{\sigma,1}=0$ locally for $E=E_*(k_y,d,m_0)$ and substitute it into
the second condition to obtain $\rho_\sigma=F_{\sigma,2}(E_*,k_y;d,m_0)$.
Its zeros locate the BICs (Sec.~\ref{supp:sec:topology} of the Supplementary Material).
At the merger, $\rho_\sigma=\partial_{k_y}\rho_\sigma=0$.
Expanding to quadratic order in momentum and first order in
length or mass gives the square-root separation:
\begin{equation}
\Delta k_y\simeq2\sqrt{\frac{|\delta|}{|b|}},\qquad
\delta=a(d-d_c)+c(m_0-m_{0,c}).
\label{eq:coalescence}
\end{equation}
Here $a$ and $c$ describe the first-order change of the residual
with length and mass, and $b$ is its quadratic momentum coefficient
at the critical point. We consider $b\ne0$ and $(a,c)\ne(0,0)$.
The critical values form a local curve in the $(d,m_0)$ plane.
For paths crossing this curve transversely, the pair exists on
the side with $\delta b<0$ to leading order. A length or mass sweep
gives the square-root approach when its corresponding linear
coefficient is nonzero.

In Fig.~\ref{fig:coalescence}(c), the opposite-charge BICs exist
for $m_0>m_{0,c}$ and approach one another as $m_0$ decreases.\
The black curves show the local square-root behaviour.
A square-root displacement of BIC positions under structural
tuning also occurs in photonic crystal slabs~\cite{ZhangLattice2026}.
Here it describes the approach of two matching zeros in momentum.
The linewidth analysis below assumes that the continued resonance
pole remains simple. This differs from exceptional BICs, where
resonance eigenvalues and their eigenvectors
coalesce~\cite{CanosValero2025Exceptional}. An acoustic realization
achieves that spectral degeneracy by balancing intermode coupling
against intrinsic-loss contrast, while symmetry suppresses radiation
into the waveguide~\cite{Wu2026Exceptional}.

To connect the loss of exact matching to electron escape, we follow
the outgoing resonance pole $z=E_{\rm r}-i\Gamma/2$, with lifetime
$\tau=\hbar/\Gamma$ for $\Gamma>0$. Away from thresholds and pole
degeneracies, a smooth decay amplitude vanishing on the BIC set
is locally proportional to $\rho_\sigma$. If the leading coupling
coefficient is nonzero, the linewidth is proportional to its square.
The outgoing matching condition and this relation are given in Sec.~\ref{supp:sec:topology}
of Supplementary Material. Using the same local expansion as for the pair separation we obtain,
\begin{equation}
\Gamma\simeq C_c
\left[\delta+b(k_y-k_{y,c})^2\right]^2,
\label{eq:unifiedwidth}
\end{equation}
where $C_c>0$ includes the pole normalization and coupling to the
single open channel of the selected mirror sector, evaluated at
the critical point.

Near either separated BIC, the residual varies linearly with
momentum displacement, giving the usual quadratic linewidth opening.
At the critical parameters, its leading variation is quadratic,
and the linewidth instead obeys
\begin{equation}
\begin{split}
\Gamma&=C_c b^2(k_y-k_{y,c})^4
+o\!\left((k_y-k_{y,c})^4\right),
\end{split}
\label{eq:quartic}
\end{equation}
at $d_c$ and $m_{0,c}$. This law concerns variation along the resonance dispersion at fixed critical parameters. Along a transverse path on the side with $\delta b>0$, there is no
nearby BIC. The local minimum linewidth is $\Gamma_{\min}\simeq
C_c\delta^2$ to leading order and vanishes quadratically on approaching
the critical curve. Figure~\ref{fig:coalescence}(d) shows these three regimes: a finite minimum below the critical mass (green), quartic opening at the critical point (black), and quadratic opening around each BIC above it (orange). The square-root separation and quartic opening thus follow from the same quadratic matching zero, with the linewidth additionally requiring regular, nonzero leading coupling.

Interestingly, the suppression of electronic escape found here has a parallel in photonic systems, where BIC merging can cancel lower-order
terms in the coupling to the continuum. Mergers in gratings and
geometrical parameter spaces can produce sixth-power suppression
of radiative losses~\cite{Bulgakov2023Super,Yu2024Merging}, and
experiments in photonic crystal slabs have demonstrated enhanced
lifetimes of nearby resonances~\cite{Jin2019}. The quartic power
here follows from the quadratic matching zero: winding constrains
the disappearance of exact confinement, while the order of the
coupling zero determines how rapidly escape resumes away from it.

\textit{Unequal masses and multilayers.---}The general confinement condition also
allows us to examine junctions without mirror symmetry.
For $m_L\ne m_R$, mirror factorization generally fails, while
the antiunitary symmetry and Eq.~\eqref{eq:general} remain valid.
There are then three real matching conditions. At fixed $k_y$,
if these conditions respond independently to changes in
$(E,d_L,d_R)$ at a BIC, small mass changes can be compensated by
adjusting its energy and the two domain widths. Confinement can
therefore persist for unequal masses, with additional tuning,
although the mirror-sector winding no longer applies (See Sec.~\ref{supp:sec:profiles} of Supplementary Material).

The same matching construction extends to other graphene
multilayers. Rhombohedral trilayers provide a natural setting
because reversals of the interlayer potential support valley
interface states~\cite{Jung2011MultilayerKinks}. Related interface
states and resonances in band continua also occur at ABC--CBA
stacking boundaries in gated trilayers~\cite{Jaskolski2020Trilayer}.
The additional propagating and evanescent channels require a
multiband description~\cite{VanDuppen2013Trilayer}. When several
decaying modes are available, confinement requires matching
linear combinations of all these modes across the junction,
rather than two individual spinors. The generalized condition
is given in Sec.~\ref{supp:sec:profiles} of Supplementary Material. It provides a test for exact confinement
in these systems, while the existence of BICs and their mirror
reduction and winding structure depend on the chosen stacking,
potential profile, and lead channels.

\textit{Experimental implications.---}Our results suggest an
electrically tunable coupling between a valley interface channel
and conducting reservoirs. Gate-defined bilayer channels, valves,
and beam splitters have already been realized~\cite{Li2016,Li2018}.
At fixed geometry, varying equal and opposite interlayer biases
while compensating the mean potential provides a route to tuning
$m_0$ through the critical point. Smooth mirror-symmetric gate
profiles admit the same sector construction, with the BIC positions
determined for the actual profile. A related control of propagation
through leakage has been demonstrated in terahertz valley photonic
crystals, where geometric modulation tunes radiation coupling and
the distance traveled by interface modes~\cite{Wei2026ValleyQBIC}.
In our junction, the corresponding escape-limited propagation
length for a narrow wave packet on a weakly leaking resonance is
$\ell_{\rm esc}\simeq\hbar|v_y|/\Gamma$, where
$v_y=\hbar^{-1}\partial E_{\rm r}/\partial k_y$.
Gate control of the escape linewidth therefore provides a route
to controlling propagation along the wall.

A direct test would combine the evolution of transport resonances
with a probe of localization near the wall.\ Measurements on
bilayer point junctions show that ordinary tunneling can contribute
alongside valley-selective
current~\cite{Davydov2024PointJunction}.\ The signature sought
here is therefore the collapse of a wall-state transmission
resonance together with persistent localization near the wall.\
Momentum-selective injection or spectroscopy is needed to resolve
the matching zeros, since ordinary conductance averages over
$k_y$.\ Testing Eq.~\eqref{eq:quartic} requires tracking the
linewidth along the resonance dispersion at the critical controls.\
For finite group velocity, this law gives
$\ell_{\rm esc}\propto|k_y-k_{y,c}|^{-4}$, suggesting reduced
leakage over a narrow momentum distribution.\ A weak local probe
or controlled detuning can access a state decoupled from the
transport channel.

Propagating surface and hinge BICs have also been observed in
phononic crystals through separability~\cite{Yin2025Hierarchy}.
Such experiments motivate optical or acoustic analogues with the
same symmetry and matching structure as the present junction,
where the predicted coalescence could be tested through
momentum-resolved scattering and spatial field measurements.
The quadratic coupling zero suggests long-lived guided resonances
with controllable outcoupling, as motivated by photonic
lifetime-engineering experiments~\cite{Jin2019}.

\textit{Conclusions.---}We have established topological BICs in
bilayer graphene mass-wall junctions, showing that a transmission
resonance can disappear while its electronic state remains
exactly confined. In symmetric junctions, their topological character is encoded in the integer winding charges of the matching zeros, which
constrain the creation and loss of confinement without changing
the valley topology supporting the wall states.
The square-root approach of opposite charges and, for regular
continuum coupling, the quartic linewidth opening follow from
the same quadratic matching zero, connecting the evolution
of topological BICs to electronic lifetimes.
The general confinement criterion also accommodates unequal
masses through additional tuning and provides a starting point
for identifying BICs in other graphene multilayers.
These findings motivate experimental realization in gate-defined
graphene junctions and optical or acoustic analogues with
equivalent symmetry and channel structure, where confinement
and escape could be controlled through geometry and external
tuning.

{\it Acknowledgments.\textendash} This research is supported by the National Natural Science Foundation of China (NSFC) grant No.\ 12547103.

\bibliographystyle{apsrev4-2}
\bibliography{references}

\onecolumngrid
\clearpage

\setcounter{equation}{0}
\setcounter{figure}{0}
\setcounter{table}{0}
\setcounter{page}{1}
\setcounter{section}{0}
\setcounter{subsection}{0}

\renewcommand{\thesection}{S\arabic{section}}
\renewcommand{\theequation}{S\arabic{equation}}
\renewcommand{\thetable}{S\arabic{table}}
\renewcommand{\thefigure}{S\arabic{figure}}
\setcounter{secnumdepth}{3}
\renewcommand{\thesubsection}{\Alph{subsection}}
\renewcommand{\thesubsubsection}{\thesubsection.\arabic{subsubsection}}

\begin{center}
\Large Supplementary Information for\\
\lq\lq Topological Bound States in the Continuum\rq\rq
\end{center}
\phantomsection
\label{sec:supplement}
\begin{center}
\normalsize Dan-Na Liu, J. Villavicencio and Pierre A. Pantale\'on
\end{center}

Here we define the main equations and analytical conditions used in the main text within our minimal four-band model.

\FloatBarrier
\section{General Model for decaying solutions}
\label{supp:sec:model}
We consider a propagation in the x-direction and translation invariance along y. The wavefunction is given by $\Psi(x,y)=e^{ik_y y}\Phi(x)$. In the basis $(A_1,B_1,B_2,A_2)$ the low energy Hamiltonian is,
\begin{equation}
H_{k_y}=
\begin{pmatrix}
V+m&\hv\pi&\gamma_1&0\\
\hv\pi^\dagger&V+m&0&0\\
\gamma_1&0&V-m&\hv\pi^\dagger\\
0&0&\hv\pi&V-m
\end{pmatrix},
\qquad \pi=-i\partial_x+ik_y.
\label{supp:eq:H}
\end{equation}
The pristine leads have $V=m=0$. In the region $0<x<L$, the sharp junction has $V=V_0$, mass $-m_L$ over width $d_L$, and mass $+m_R$ over
width $d_R$, where $L=d_L+d_R$. The particular symmetric case has $m_L=m_R=m_0$ and $d_L=d_R=d=L/2$. Uniform-region modes follow by using $\Phi(x)=\bm u(k_x)e^{ik_xx}$ into Eq.~\eqref{supp:eq:H}. With $e_m=E-V-m$ and $e_p=E-V+m$, their longitudinal wave numbers satisfy 
\begin{equation}
\bigl[e_m^2-(\hv)^2(k_x^2+k_y^2)\bigr]
\bigl[e_p^2-(\hv)^2(k_x^2+k_y^2)\bigr]
-\gamma_1^2e_me_p=0.
\label{supp:eq:dispersion}
\end{equation}
This equation allows real, imaginary, and complex wavenumbers $k_x$. To identify the propagating and evanescent channels we write 
\begin{equation}
0<E<\gamma_1,\qquad
k_{\rm p}^2=\frac{E(E+\gamma_1)}{(\hv)^2}-k_y^2>0,\qquad
\kappa^2=k_y^2+\frac{E(\gamma_1-E)}{(\hv)^2}>0.
\label{supp:eq:channels}
\end{equation}
We take $k_{\rm p},\kappa>0$. Each lead then has one propagating and one evanescent channel. To solve the confinement problem we write the spinors decaying toward the left and right infinities as
\begin{equation}
\bm u_L=
\begin{pmatrix}
1\\-i\hv(\kappa+k_y)/E\\1\\-i\hv(\kappa-k_y)/E
\end{pmatrix},
\qquad
\bm u_R=
\begin{pmatrix}
1\\i\hv(\kappa-k_y)/E\\1\\i\hv(\kappa+k_y)/E
\end{pmatrix},
\label{supp:eq:tails}
\end{equation}
with their spatial factors given by $e^{\kappa x}$ and
$e^{-\kappa(x-L)}$, respectively. Equation~\eqref{supp:eq:tails} with the conditions in Eq.~\eqref{supp:eq:channels} allow us to define the transport properties and in particular the confinement conditions. 

\FloatBarrier
\section{General BIC condition and mirror reduction}
\label{supp:sec:matching}

We define the propagator of Eq.~\eqref{supp:eq:H} through $\Phi(x_2)=\mathsf P(x_2,x_1)\Phi(x_1)$. Matching requires continuity of all four spinor components at every finite potential step. At any $x_*\in[0,L]$, we have
\begin{equation}
\begin{split}
P_L&=\mathsf P(x_*,0),\qquad P_R=\mathsf P(L,x_*),\\
\bm v_L&=P_L\bm u_L,\qquad
\bm v_R=P_R^{-1}\bm u_R.
\end{split}
\label{supp:eq:propagated}
\end{equation}
where the inverse propagates the right tail backward. Since propagation is invertible, both vectors in the second row are nonzero. Because the vectors in Eq.~\eqref{supp:eq:propagated} represent decaying solutions in the corresponding side of the central lead then a BIC exists exactly when they are proportional, or equivalently when
\begin{equation}
\mathcal D_{\rm BIC}
=\norm{\bm v_L}^2\norm{\bm v_R}^2
-|\bm v_L^\dagger\bm v_R|^2=0.
\label{supp:eq:general}
\end{equation}
A BIC exists precisely when the unique decaying modes in the two leads match continuously through the junction, forming an eigenfunction with finite norm in $x$. In the interior it is
$\Phi_B(x)=\mathcal N\mathsf P(x,0)\bm u_L$, where $\mathcal N$
normalizes the complete state, including both semi-infinite lead
tails, note that depending on the form of the propagator, internal propagating components are allowed. The BIC condition is independent of the matching point and the normalization chosen for the decaying lead spinors. 

At fixed $k_y$, the Hamiltonian commutes with the antiunitary operator
\begin{equation}
\Theta=Z\mathcal K,\qquad
Z=\diag(1,-1,1,-1),\qquad \Theta^2=1,
\label{supp:eq:Theta}
\end{equation}
where $\mathcal K$ denotes complex conjugation. Its action is
$\Theta\Phi=Z\Phi^*$: it conjugates scalar coefficients,
$\Theta(c\Phi)=c^*\Theta\Phi$, while preserving the norm.
The lead spinors satisfy $Z\bm u_{L,R}^*=\bm u_{L,R}$,
and propagation at real energy preserves this relation.
Thus their $A_1,B_2$ components are real and
their $B_1,A_2$ components are imaginary in the chosen normalization.
Proportionality of two real four-component vectors imposes three
independent conditions after their relative amplitude is removed. We note that this symmetry acts within one valley at fixed
$k_y$ and is distinct from physical time reversal.
For potential profiles of the form $V(L-x)=V(x)$ and $m(L-x)=-m(x)$, the Hamiltonian also commutes with
\begin{equation}
(\mathcal M\Phi)(x)=\Xi\Phi(L-x),\qquad
\Xi=
\begin{pmatrix}
0&0&1&0\\0&0&0&-1\\1&0&0&0\\0&-1&0&0
\end{pmatrix}.
\label{supp:eq:mirror}
\end{equation}
Since $\mathcal M^2=1$, we can define a BIC mirror parity as $\sigma=\pm1$. The half-junction condition is given by
\begin{equation}
\bm r_\sigma=\mathsf B_\sigma P_L\bm u_L=0,\qquad
\mathsf B_\sigma=
\begin{pmatrix}1&0&-\sigma&0\\0&1&0&\sigma\end{pmatrix},
\qquad P_L=\mathsf P(d,0).
\label{supp:eq:sector}
\end{equation}
where the right half follows from $\Phi(x)=\sigma\Xi\Phi(L-x)$.
At $x_*=d$, reflection gives $\bm v_R=\Xi\bm v_L$.
Using $\mathsf B_\sigma^\dagger\mathsf B_\sigma=I_4-\sigma\Xi$
in Eq.~\eqref{supp:eq:general}, the general BIC condition factorizes exactly:
\begin{equation}
\mathcal D_{\rm BIC}
=\norm{\mathsf B_+\bm v_L}^2\norm{\mathsf B_-\bm v_L}^2.
\label{supp:eq:factorization}
\end{equation}
The two factors cannot vanish simultaneously, since satisfying both mirror-sector boundary conditions would force the entire wavefunction to vanish. Interestingly, the antiunitary symmetry makes $r_{\sigma,1}$ real and $r_{\sigma,2}$ imaginary. We collect the two real matching conditions in
$\bm F_\sigma$ and define the complex function $f_\sigma$: 
\begin{equation}
\bm F_\sigma=
\begin{pmatrix}
\operatorname{Re}r_{\sigma,1}\\
\operatorname{Im}r_{\sigma,2}
\end{pmatrix},
\qquad
f_\sigma=F_{\sigma,1}+iF_{\sigma,2},\qquad
\mathcal D_{\rm BIC}=|f_+|^2|f_-|^2.
\label{supp:eq:map}
\end{equation}
Because $\sigma$ defines the parity, then each mirror sector contains an open channel formed by combining incidence from the two leads. The confinement is then obtained when both real mirror sector equations vanish.

\FloatBarrier
\section{Spatial spectral maps.}
\label{supp:sec:spectra}
The same mode-matching construction gives the retarded Green function
$\mathcal G^R(x,x';E+i\eta,k_y)$, which satisfies
$[E+i\eta-H_{k_y}]\mathcal G^R(x,x';E+i\eta,k_y)
=\delta(x-x')I_4$.
For $x\ne x'$, its columns obey the homogeneous equation and are
matched continuously at the contacts and mass wall. The lead
conditions are continued from outgoing propagating modes and
decaying evanescent modes; the delta source fixes the discontinuity
at $x=x'$. At coincident positions, the diagonal components give
the local spectral densities on the two layers:
\begin{equation}
\begin{split}
D_{1G}^\eta&=-\frac{1}{\pi}\operatorname{Im}
(\mathcal G^R_{11}+\mathcal G^R_{22}),\qquad
D_{2G}^\eta=-\frac{1}{\pi}\operatorname{Im}
(\mathcal G^R_{33}+\mathcal G^R_{44}),\\
D_G^\eta&=D_{1G}^\eta+D_{2G}^\eta,\qquad
P_L^\eta=\frac{D_{1G}^\eta-D_{2G}^\eta}{D_G^\eta}.
\end{split}
\label{supp:eq:spectraldensities}
\end{equation}
Here all Green-function components are evaluated at
$(x,x;E+i\eta,k_y)$ with $\eta>0$.
Thus $D_G^\eta$ maps the spatial distribution of spectral weight,
while $P_L^\eta$ measures its layer imbalance.
The spectral broadening $\eta$ is distinct from the escape
linewidth $\Gamma$, which vanishes at a BIC.

To distinguish the bound state from the background at the same
position and energy, we separate the pole of an isolated BIC at
$E_B$:
\begin{equation}
\mathcal G^R(x,x';z)
=\frac{\Phi_B(x)\Phi_B^\dagger(x')}{z-E_B}
+\mathcal G_{\rm reg}^R(x,x';z),\qquad z=E+i\eta.
\label{supp:eq:BICspectralpole}
\end{equation}
Here $\mathcal G_{\rm reg}^R$ is the remainder of the retarded
Green function after subtracting the selected BIC contribution.
It retains the continuum background but contains no pole
associated with this BIC at $E_B$. Also, $\Phi_B$ is normalized on the entire $x$ axis, including the
lead tails, as specified in Sec.~\ref{supp:sec:matching}. With
$\Lambda=\diag(1,1,-1,-1)$, define
$\rho_B=\Phi_B^\dagger\Phi_B$ and
$P_B=\Phi_B^\dagger\Lambda\Phi_B/\rho_B$.
The BIC contributes
$D_B^\eta=\rho_B\eta/\{\pi[(E-E_B)^2+\eta^2]\}$ to the
local spectral density. Applying Eq.~\eqref{supp:eq:spectraldensities}
to $\mathcal G_{\rm reg}^R$ defines $D_{\rm reg}^\eta$ and
$P_{\rm reg}^\eta$, so
\begin{equation}
D_G^\eta=D_B^\eta+D_{\rm reg}^\eta,\qquad
P_L^\eta=
\frac{D_B^\eta P_B+D_{\rm reg}^\eta P_{\rm reg}^\eta}
{D_B^\eta+D_{\rm reg}^\eta}.
\label{supp:eq:polarizationdecomposition}
\end{equation}
Polarizations are defined where their corresponding densities
are nonzero. Thus a BIC reverses the local background contrast
when $P_B$ and $P_{\rm reg}^\eta$ have opposite signs and
$D_B^\eta|P_B|>D_{\rm reg}^\eta|P_{\rm reg}^\eta|$.
The zero-polarization contour has equal layer spectral densities,
not a vanishing total density. This comparison is local: spatial
averages can have the same sign even when the two contributions
favor opposite layers in part of the junction. Note that neither
the polarization contrast nor such an average replaces the
confinement criterion in Eq.~\eqref{supp:eq:general}.

\FloatBarrier
\section{Unequal masses, smooth profiles, and multilayers}
\label{supp:sec:profiles}

For unequal masses or widths, the BIC condition follows from
matching the decaying lead modes through the full junction using
Eq.~\eqref{supp:eq:general}. Locally, the three real matching
conditions can be written by dividing each propagated tail by the same
nonzero component and equating the other three components. At fixed
$k_y$, if the Jacobian of these conditions with respect to
$(E,d_L,d_R)$ is nonsingular at a BIC, the implicit function theorem
gives a nearby BIC for small changes in the masses, with adjusted
energy and widths.

Smooth mass walls and contacts are treated
by dividing the chosen profiles $V(x)$ and $m(x)$ into locally
uniform slabs~\cite{LiuMode2026}. Continuity of the four-component
spinor gives the full propagator
\begin{equation}
\begin{split}
\mathsf P(L,0)&=\mathsf P_N\cdots\mathsf P_1,\\
\mathsf P_j&=\Omega_j
\operatorname{diag}_{\alpha=1,\ldots,4}
\!\left(e^{iq_{j,\alpha}\Delta x_j}\right)
\Omega_j^{-1},
\end{split}
\end{equation}
where $\Delta x_j$ is the slab width, and the columns of
$\Omega_j$ are the four local eigenvectors of
Eq.~\eqref{supp:eq:H}, with longitudinal wave numbers $q_{j,\alpha}$
evaluated at $V_j$ and $m_j$. This modal form applies
when $\Omega_j$ is invertible. For any slab, including a defective
mode degeneracy, use $\mathsf P_j=\exp(\mathsf A_j\Delta x_j)$, where
$\partial_x\Phi=\mathsf A_j\Phi$ follows from
Eq.~\eqref{supp:eq:H}. Writing the slab Hamiltonian as
$H_{k_y}=-i\hv\mathsf J\partial_x+H_{0,j}$, with $\mathsf J$ exchanging
the two sublattice components within each layer, gives
$\mathsf A_j=i\mathsf J(EI_4-H_{0,j})/(\hv)$. This construction retains all
propagating and evanescent components and converges to the
smooth-profile propagator as the slab widths vanish. The same
propagation determines the transmission amplitudes and the
decaying-tail matching in Eq.~\eqref{supp:eq:general}. The reduction
to mirror sectors remains exact whenever $V(L-x)=V(x)$ and
$m(L-x)=-m(x)$. 

\subsection{Extension to multilayer graphene}

The same confinement construction applies when the leads contain
several independent decaying modes, as can occur in multilayer
graphene. Valley interface states in rhombohedral multilayers
and multiband transport in trilayers provide relevant
settings~\cite{Jung2011MultilayerKinks,VanDuppen2013Trilayer}. We consider an $N$-layer model with $2N$ spinor components and an
invertible propagator and let the columns of $U_L$ and $U_R$ contain the contact spinors of all modes decaying toward the left and right
infinities. Their numbers are $r_L$ and $r_R$, respectively.
At fixed real $(E,k_y)$ away from lead thresholds, a confined
solution must satisfy
\begin{equation}
\mathsf P(L,0)U_L\bm a=U_R\bm b,
\qquad (\bm a,\bm b)\ne0,
\label{supp:eq:multilayermatching}
\end{equation}
where $\bm a$ and $\bm b$ are the coefficients of the decaying
lead modes. Thus confinement occurs when the matching matrix
has linearly dependent columns:
\begin{equation}
\operatorname{rank}
\begin{pmatrix}\mathsf P(L,0)U_L&-U_R\end{pmatrix}
<r_L+r_R.
\label{supp:eq:multilayerrank}
\end{equation}
The columns of each $U_{L,R}$ are linearly independent, so a
nonzero coefficient solution gives nonzero tails on both sides.
Equation~\eqref{supp:eq:multilayermatching} joins them into a continuous
eigenfunction with finite norm in $x$. It is a BIC when the
leads also support propagating states at the same $(E,k_y)$.

For $r_L=r_R=1$, this condition reduces to proportionality of
the two propagated tails and hence to Eq.~\eqref{supp:eq:general}.
For a trilayer, the contact spinors have six components, and
the decaying-mode count follows from the chosen lead spectrum.
The criterion determines whether a confined state exists; it
does not assume that every valley interface mode becomes a BIC.
Mirror reduction can be used when the full multilayer model
has an appropriate reflection symmetry. Its boundary matrices
and reduced matching equations must then be obtained for that
model before assigning winding charges.

\FloatBarrier
\section{Winding, coalescence, and linewidth scaling}
\label{supp:sec:topology}

Within each mirror sector, BICs are zeros of the matching
function $f_\sigma$ in the $(E,k_y)$ plane. The phase around
these zeros defines a winding charge that constrains how
they can merge as the junction length or mass is varied.
We first characterize this winding, then examine the matching
conditions near a particular point and their relation to the resonance
linewidth.

The winding charge counts the net change of the phase of
$f_\sigma$ around a closed contour $\mathcal C$:
\begin{equation}
q_\sigma=\frac{1}{2\pi}\oint_{\mathcal C}d\arg f_\sigma.
\label{supp:eq:winding}
\end{equation}
The contour is traversed counterclockwise in the ordered
$(E,k_y)$ plane and must contain no zeros on its boundary.
A phase increase or decrease of $2\pi$ gives charge $+1$
or $-1$, respectively. When $\mathcal C$ encloses a single simple zero, the charge can be calculated directly from the local matching equations:
\begin{equation}
q_\sigma=\operatorname{sgn}
\det\frac{\partial(F_{\sigma,1},F_{\sigma,2})}
{\partial(E,k_y)}.
\label{supp:eq:windinglocal}
\end{equation}
The derivatives are evaluated at the BIC. At a simple zero,
this determinant is nonzero, and its sign determines the
direction of the phase winding.
The component and coordinate orders fix its sign.
Winding is conserved while the contour remains zero-free and the
matching map, symmetries, and channel structure remain regular. Changing the junction length or mass moves the BIC zeros in the
$(E,k_y)$ plane. Two zeros may approach and meet at a critical
parameter value, forming a merged BIC; beyond this value, the
pair may be absent. To determine when this disappearance is
allowed, consider a contour enclosing both zeros. Its total
winding remains unchanged as long as no zero crosses the contour
and the matching equations and lead-channel regime remain regular.
The pair can therefore disappear only if its charges sum to zero.

This conservation applies separately to each mirror sector.
Opposite charges within the same sector allow pair annihilation,
but do not guarantee that a chosen parameter tuning brings the
zeros together. BICs from opposite mirror sectors cannot meet
at the same $(E,k_y)$ for the same junction parameters, because
the nonzero decaying lead solution cannot satisfy both sector
boundary conditions. The winding defined here characterizes
exact decoupling from the continuum and is distinct from the
valley Chern numbers of the gapped domains.

\subsection{Merging Conditions}
To describe how two BICs approach and merge, we combine the
two real BIC conditions $F_{\sigma,1}=F_{\sigma,2}=0$ in
Eq.~\eqref{supp:eq:map} into a single condition for $k_y$ and
the control parameters $d$ and $m_0$. When
$\partial_EF_{\sigma,1}\ne0$, the first condition determines
the energy locally as $E=E_*(k_y,d,m_0)$. Substituting this
energy into the second condition then gives
\begin{equation}
\rho_\sigma(k_y,d,m_0)
=F_{\sigma,2}(E_*(k_y,d,m_0),k_y;d,m_0).
\label{supp:eq:residual}
\end{equation}

Its zeros satisfy both matching conditions and therefore
locate the BICs. At the critical point, the two roots coincide,
so $\rho_\sigma=\partial_{k_y}\rho_\sigma=0$.
Expanding around the critical point
$(k_{y,c},d_c,m_{0,c})$ gives
\begin{equation}
\begin{split}
\rho_\sigma
={}&a(d-d_c)+c(m_0-m_{0,c})+b(k_y-k_{y,c})^2+\cdots .
\end{split}
\label{supp:eq:fold}
\end{equation}
Here $a=\partial_d\rho_\sigma$, $c=\partial_{m_0}\rho_\sigma$, and
$b=\tfrac12\partial_{k_y}^2\rho_\sigma$ are evaluated at the critical point.
We consider the case $b\ne0$ and $(a,c)\ne(0,0)$.
The equations $\rho_\sigma=\partial_{k_y}\rho_\sigma=0$ define a
local critical curve in the $(d,m_0)$ plane. Write
$\delta=a(d-d_c)+c(m_0-m_{0,c})$. For a path crossing the curve
transversely, the side with $\delta b<0$ contains two nearby BICs,
with leading-order momentum separation
\begin{equation}
\Delta k_y\simeq2\sqrt{\frac{|\delta|}{|b|}}.
\label{supp:eq:coalescence}
\end{equation}
On a tangent path, $\delta$ vanishes to first order in the path
parameter and higher-order terms can determine the side of the curve.
The two BICs approach the critical point with opposite winding
charges. To see this, note that the two roots of
Eq.~\eqref{supp:eq:fold} have opposite slopes
$\partial_{k_y}\rho_\sigma$. Along $F_{\sigma,1}=0$,
\[
\det\frac{\partial(F_{\sigma,1},F_{\sigma,2})}
{\partial(E,k_y)}
=(\partial_EF_{\sigma,1})\partial_{k_y}\rho_\sigma.
\]
Since $\partial_EF_{\sigma,1}$ remains nonzero and has the same
sign near both roots, their charges are opposite. At the
critical point, the two roots coincide and their total winding
is zero, but the matching conditions still hold: the resulting
state remains a BIC. The pair is absent on the other side
of a transverse crossing.

\subsection{Resonance linewidth}
To determine how electrons escape when the BIC condition is
no longer satisfied, we extend the matching problem to
resonances. A resonance permits outgoing propagating waves
in the leads, together with the evanescent components.
The left-outgoing propagating spinor at real energy is
\begin{equation}
\bm u_{{\rm out},L}
=\left(1,-\frac{\hv(k_{\rm p}+ik_y)}{E},
-1,\frac{\hv(k_{\rm p}-ik_y)}{E}\right)^T.
\label{supp:eq:outgoingspinor}
\end{equation}
The resonance energies are obtained by continuing these lead
modes to complex energy and imposing the same mirror-sector
boundary conditions:
\begin{equation}
\mathcal D_\sigma^{\rm out}(z,k_y;d,m_0)
=\det\left[
\mathsf B_\sigma P_L(z,k_y;d,m_0)
\begin{pmatrix}
\bm u_{{\rm out},L}(z,k_y)&\bm u_L(z,k_y)
\end{pmatrix}
\right]=0,
\label{supp:eq:sectorpole}
\end{equation}
where $z=E_{\rm r}-i\Gamma/2$. Here $E_{\rm r}$ is the resonance energy and $\Gamma$ is its
escape linewidth, with lifetime $\tau=\hbar/\Gamma$ for
$\Gamma>0$. At a BIC, the outgoing propagating amplitude
vanishes and $\Gamma=0$. The bound state remains even when
its transmission resonance collapses, because its singularity
can be removable in the propagating scattering amplitude.
A transmission zero alone therefore does not establish a BIC;
the decaying-tail condition must also be satisfied.

The distinction between layer contrast and confinement is also
explicit in probability conservation. For an outgoing solution
$\Phi_z$ at $z=E_{\rm r}-i\Gamma/2$, define the probability
current
\begin{equation}
\begin{split}
j_x(x)&=2v_F\operatorname{Re}
\left[\Phi_{z,A_1}^*\Phi_{z,B_1}
+\Phi_{z,B_2}^*\Phi_{z,A_2}\right],\\
\hbar\bigl[j_x(x_b)-j_x(x_a)\bigr]
&=\Gamma\int_{x_a}^{x_b}\Phi_z^\dagger(x)\Phi_z(x)\,dx.
\end{split}
\label{supp:eq:escapeflux}
\end{equation}
This identity follows directly from Eq.~\eqref{supp:eq:H} and holds
on any finite interval $[x_a,x_b]$; an outgoing resonance is
not normalized over the infinite leads. It relates the net
outgoing probability flux to the decay of the weight in that
interval. The current depends on component coherences, whereas
the layer polarization compares populations. At a BIC, the
propagating contact amplitudes and $\Gamma$ vanish. A reversal
of layer polarization imposes neither condition. The broadening
$\eta$ in Sec.~\ref{supp:sec:spectra} remains a spectral resolution
parameter and is not an escape rate.

The connection between coalescence and lifetime follows from
how the escape amplitude grows away from exact matching.
Consider a simple outgoing pole away from thresholds and
other pole degeneracies, and let $g_\sigma$ be its smooth
decay amplitude along the resonance dispersion. This amplitude
vanishes wherever $\rho_\sigma=0$. Since $\rho_\sigma$ has
a nonzero derivative with respect to at least one control
parameter, a smooth amplitude vanishing on this set can be
written locally as
\begin{equation}
g_\sigma=u_\sigma\rho_\sigma,\qquad
\Gamma=C\rho_\sigma^2+o(\rho_\sigma^2),\qquad C>0,
\label{supp:eq:widthresidual}
\end{equation}
provided the smooth coefficient $u_\sigma$ is nonzero at the
critical point. The linewidth is thus proportional, to leading
order, to the squared matching residual. The coefficient $C$
includes the pole normalization and coupling to the single
open channel of the selected mirror sector. Here $E_*$
is the energy obtained from the first real matching condition;
away from a BIC, it need not equal the resonance energy
$E_{\rm r}$.

Substituting the local expansion in Eq.~\eqref{supp:eq:fold} gives
\begin{equation}
\Gamma\simeq C_c[\delta+b(k_y-k_{y,c})^2]^2,
\label{supp:eq:unifiedwidth}
\end{equation}
where $C_c$ is evaluated at the critical point.
For a transverse crossing on the side with $\delta b>0$, the local
minimum linewidth is $\Gamma_{\min}\simeq C_c\delta^2$ to leading order.
Near either separated BIC, the residual varies linearly with
momentum displacement, so the linewidth opens quadratically.
At the critical controls, the linear variation vanishes and
the residual begins at second order, giving
\begin{equation}
\Gamma=C_cb^2(k_y-k_{y,c})^4+
o\!\left((k_y-k_{y,c})^4\right),\qquad
(d,m_0)=(d_c,m_{0,c}).
\label{supp:eq:quartic}
\end{equation}
Thus the quartic linewidth follows from the quadratic matching
zero together with a smooth, nonzero leading coupling
coefficient.

\end{document}